\magnification=1200
\baselineskip=14pt

\def\P{\Phi}
\def\L{\Lambda}
\def\a{\alpha}
\def\b{\beta}

\def\d{\delta}
\def\l{\lambda}
\def\N{{\cal N}}
\def\G{{\cal G}}
\def\F{{\cal F}}
\def\half{{1\over 2}}
\def\tr{{\rm tr}}
\def\p{\partial}

\rightline{UCLA/99/TEP/21}
\rightline{Columbia/99/Math}

\bigskip

\centerline{{\bf SEIBERG-WITTEN THEORY AND CALOGERO-MOSER SYSTEMS}
\footnote*{Research supported in part by the 
National
Science Foundation under grants PHY-95-31023 and DMS-98-00783.}}

\bigskip
\bigskip

\centerline{\bf Eric D'Hoker${}^1$ and D.H. Phong${}^2$}

\bigskip

\centerline{${}^1$ Department of Physics}
\centerline{University of California, Los Angeles, CA 90095}

\medskip

\centerline{${}^2$ Department of Mathematics}
\centerline{Columbia University, New York, NY 10027}

\bigskip

\centerline{\bf Abstract}

\bigskip

We present a brief account of a series of recent results on 
twisted and untwisted elliptic 
Calogero-Moser systems, and on their fundamental role in the 
Seiberg-Witten solution of gauge theories with one massive hypermultiplet 
in the adjoint representation of an arbitrary gauge algebra $\G$. 

\bigskip
\bigskip

\centerline{{\it Contribution to the Proceedings of the}}
\centerline{{\it ``Workshop on Gauge Theory and Integrable Models", Kyoto,
January 1999,}} 

\vfill\break

\centerline{\bf I. INTRODUCTION}

\bigskip

Over the past few years, a tremendous amount of progress has been 
made in the understanding of strongly coupled supersymmetric
gauge and string theories. These advances were driven in large 
part by the Seiberg-Witten solution of $\N=2$ supersymmetric 
Yang-Mills theory for $SU(2)$ gauge group [1]. 
     
\medskip

Of central interest to many of these developments
is the 4-dimensional supersymmetric Yang-Mills theory
with maximal supersymmetry, $\N=4$, and with arbitrary gauge 
algebra $\G$. In the present paper, we shall consider a
generalization of this theory, in which a mass
term is added for part of the $\N=4$ gauge multiplet,
softly breaking the $\N=4$ symmetry to $\N=2$.  
As an $\N=2$ supersymmetric theory, the theory has a $\G$-gauge
multiplet, and a hypermultiplet in the adjoint representation 
of $\G$ with mass $m$. 
This generalized theory shares many of the properties
of the $\N=4$ theory : it has the same field content;
it is ultra-violet finite;
it has vanishing renormalization group $\beta$-function,
and it is expected to have Montonen-Olive duality symmetry.
For vanishing hypermultiplet mass $m=0$, the $\N=4$ theory
is recovered. For $m\to \infty$, the limiting theory is one of many
interesting $\N=2$ supersymmetric Yang-Mills theories.
The possibilities for, say, $\G=SU(N)$, 
include theories with any number of hypermultiplets in the
fundamental representation of $SU(N)$, or with product
gauge algebras $SU(N_1)\times SU(N_2) \times \cdots \times SU(N_p)$,
and hypermultiplets in the fundamental and 
bi-fundamental representations of these product algebras.
   
\medskip

Remarkably, the Seiberg-Witten theory for $\N=2$ supersymmetric
Yang-Mills theory for arbitrary gauge algebra $\G$ appears 
to be intimately related with the existence of certain  
classical mechanics integrable systems. This relation was
first suspected on the basis of the similarity between the Seiberg-Witten
curves and the spectral curves of certain integrable models [2].
Then, general arguments showed that Seiberg-Witten Ansatz 
naturally produces integrable structures [3]. 
For the $\N=2$ supersymmetric Yang-Mills theory with massive
hypermultiplet, the relevant integrable system appears to be
the {\it elliptic Calogero-Moser system}. For $SU(N)$ gauge group, 
Donagi and Witten [3] proposed that the spectral curves of the 
$SU(N)$ Hitchin system should play the role of the Seiberg-Witten 
curves. Krichever (in unpublished work), Gorsky and Nekrasov,
and Martinec [4]
recognized that the $SU(N)$ Hitchin system spectral curves
are identical to those of the $SU(N)$ elliptic Calogero-Moser
integrable system. 

\medskip

That the $SU(N)$ elliptic Calogero-Moser
curves (and associated Seiberg-Witten differential) do
indeed  provide the Seiberg-Witten solution for the $\N=2$
theory with one massive hypermultiplet was fully established 
by the authors in [5]. There, it was shown that
the effective prepotential $\F$ reproduces the logarithmic
singularities predicted by perturbation theory; that
$\F$ satisfies a renormalization group type equation which
determines instanton contributions to
any order; and that the prepotential in the limit of large hypermultiplet 
mass $m$ reproduces the prepotentials for $\N=2$ super Yang-Mills theory with 
any number of hypermultiplets in the fundamental representation of the gauge 
group.

\medskip

The fundamental problem in Seiberg-Witten theory
is to determine the Seiberg-Witten curves and differentials, 
corresponding to an $\N=2$ supersymmetric gauge theory with 
arbitrary gauge algebra $\G$, and a massive hypermultiplet in
an arbitrary representation $R$ of $\G$, subject to
the constraint of asymptotic freedom or conformal invariance.
With the correspondence between Seiberg-Witten 
curves and the spectral curves of classical mechanics 
integrable systems [3], this problem is equivalent to 
determining a general integrable system, associated 
with the Lie algebra $\G$ and the representation $R$.

\medskip

The $\N=2$ theory for arbitrary gauge algebra $\G$ and
with one massive hypermultiplet in the adjoint representation 
was one such outstanding case when $\G \not= SU(N)$.
Actually, as discussed previously, upon taking suitable 
limits, this theory contains a very large number of
models with smaller hypermultiplet representations $R$,
and in this sense has a universal aspect. It appeared
difficult to generalize directly the Donagi-Witten
construction of Hitchin systems
to arbitrary $\G$, and it was thus natural 
to seek this generalization directly amongst the elliptic Calogero-Moser
integrable systems.
It has been known now for a long time,
thanks to the work of Olshanetsky and Perelomov [6],
that Calogero-Moser systems can be defined for any simple
Lie algebra.
Olshanetsky and Perelomov also showed that
the Calogero-Moser systems for {\it classical} Lie
algebras were integrable,
although the existence of a spectral curve
(or Lax pair with spectral parameter) as well as the case
of exceptional Lie algebras remained open.

\medskip

The purpose of this paper is to review the resolution 
of the above problems by the following results
which were obtained in [7], [8], [9]. 

\medskip

$\bullet$ The elliptic Calogero-Moser systems defined
by an arbitrary simple Lie algebra $\G$ admit Lax
pairs with spectral parameters.

$\bullet$ The correspondence between elliptic $\G$ Calogero-Moser
systems and $\N=2$ supersymmetric $\G$ gauge theories
with matter in the adjoint representation holds directly when
the Lie algebra $\G$ is simply-laced.
When $\G$ is not simply-laced, the correspondence is with 
new integrable models, {\it the twisted elliptic Calogero-Moser systems}
introduced in [7,8]. 

$\bullet$ Twisted elliptic Calogero-Moser systems
admit a Lax pair with spectral parameter~[7].

$\bullet$ In the scaling limit $m=Mq^{-{1\over 2}\d}\rightarrow\infty$,
$M$ fixed, the twisted (respectively untwisted) elliptic $\G$
Calogero-Moser
systems tend to the Toda system for
$(\G^{(1)})^{\vee}$ (respectively $\G^{(1)}$)
for $\d={1\over h_{\G}^{\vee}}$ (respectively $\d={1\over h_{\G}}$).
Here $h_{\G}$ and $h_{\G}^{\vee}$ are the Coxeter and the dual Coxeter numbers
of $\G$ [8]. 

\medskip

The remainder of this paper is organized as follows. In \S II, we review
the set-up and basic constructions of Seiberg-Witten theory. In \S III, we 
discuss the elliptic Calogero-Moser systems, and present
the new twisted elliptic Calogero-Moser systems introduced in [7,8]. In \S IV, 
we show how these systems tend to Toda systems in certain limits. In \S V, we 
discuss their integrability properties and present their Lax pairs with 
spectral parameter. Finally, in \S VI, we discuss the Seiberg-Witten solution 
for the $\N=2$ supersymmetric Yang-Mills theories and a massive hypermultiplet 
in the adjoint representation of an arbitrary gauge algebra $\G$. A prior 
review of these results has appeared in [10], where further results 
on spin Calogero-Moser systems are also presented.

\bigskip

\centerline{\bf II. SEIBERG-WITTEN THEORY}

\medskip

The starting point for Seiberg-Witten theory is an $\N=2$ supersymmetric 
Yang-Mills theory with gauge algebra $\G$ and hypermultiplets in a 
representation $R$ of $\G$ with masses $m_j$. The microscopic Lagrangian is 
completely fixed by $\N=2$ supersymmetry in terms of the gauge coupling $g$ and 
the instanton angle $\theta$, and is given by
$$
{\cal L} = {1 \over 4 g^2} F_{\mu \nu}^a F^{\mu \nu a}
 + {\theta \over 32 \pi ^2} F_{\mu \nu}^a \tilde F^{\mu \nu a}
 + D_\mu \bar \phi D^\mu \phi + \tr [\bar \phi , \phi ]^2 + \cdots
 \eqno (2.1)
$$
where we have neglected hypermultiplet and fermion terms.

\medskip

The low energy effective theory corresponding to this model can be analyzed
by studying first the structure of the vacuum. $\N=2$ supersymmetric vacuum 
states can occur whenever the vacuum energy is exactly zero, which is achieved 
for constant scalar fields $\phi$ for which the potential energy term  
vanishes. This requires $[\bar \phi, \phi]=0$,  so that
the vacuum expectation value of $\phi$ is
a linear combination of the Cartan generators $h_j$ of the gauge algebra $\G$,
$$
< \phi > = \sum _{j=1} ^n a_j h_j
\qquad \qquad
n={\rm rank} ~\G
\eqno (2.2)
$$
Here, the complex parameters $a_j$ are usually referred to as the quantum 
moduli, or also as the quantum order parameters of the $\N=2$ vacua. 

\medskip

For generic values of the parameters $a_j$, the $\G$-gauge symmetry will be 
broken down to $U(1)^n/{\rm Weyl}(\G)$, and the low energy theory is that of 
$n$ different Coulomb fields, up to global identifications by ${\rm
Weyl}(\G)$. The low energy effective Lagrangian is invariant under $\N=2$ 
supersymmetry and thus given by 
$$
{\cal L} _{\rm effective} 
= {1 \over 4}{\rm Im} (\tau _{ij}) F_{\mu \nu} ^i F^{\mu \nu j}
+ {1 \over 4} {\rm Re} (\tau _{ij}) F_{\mu \nu} ^i \tilde F^{\mu \nu j}
+ \partial _\mu \bar \phi ^j \partial ^\mu \phi _{Dj} + {\rm fermions}
\eqno (2.3)
$$
Here, the dual gauge scalar $\phi _D$ and the gauge coupling function $\tau 
_{ij}$ are both given in terms of the prepotential $\F$
$$
\phi _{Dj} = {\partial \F(\phi) \over \partial \phi _j}
\qquad
\tau _{ij} = {\partial ^2 \F (\phi) \over \partial \phi _i \partial \phi _j}
\eqno (2.4)
$$
The form of the effective Lagrangian (2.3) is the same for any of the values
of the complex  moduli of $\N=2$ vacua, with the understanding that the
fields $\phi  _j$ take on the expectation value $<\phi _j>=a_j$. Since the
prepotential $\F(\phi)$ is a function of the fields $\phi$ only, but not of
derivatives of $\phi$, the prepotential will be completely determined by its
values on the  vacuum expectation values of the field, namely by its values on
the quantum order parameters $a_j$.

\medskip

The object of Seiberg-Witten theory is the determination of the prepotential 
$\F (a_j)$, from which the entire low energy effective action will be known. 
This is achieved by exploiting the physical conditions satisfied by $\F$ [1],

\medskip 

\item{(1)} $\F(a_j)$ is complex analytic in $a_j$ in view of $\N=2$ 
supersymmetry.

\item{(2)} The matrix ${\rm Im}\ \tau _{ij} = {\rm Im}\, \partial _i 
\partial _j 
\F$ is positive definite, since by (2.3), it coincides with the metric on the 
kinetic terms for the gauge fields $A_j$.

\item{(3)} The large $a_j$ behavior is known from perturbative quantum field 
theory calculations and asymptotic freedom, and is given by
$\F(a) \sim \sum_{\alpha}(\alpha\cdot a)^2 \ln (\alpha\cdot a)^2
-\sum_{w}(w\cdot a+m)^2\ln(w\cdot a+m)^2$,
where $\alpha$ and $w$ are respectively the roots of $\G$ and the weights
of the representation $R$.

\medskip

\noindent
As a result of (1) and (2), $\F$ 
cannot be a single-valued function of the $a_j$. For if it were, ${\rm Im} \ 
\tau _{ij}$ would be both harmonic and bounded from below, implying 
that it must be independent of $a_j$. But from (3) we know that $\tau _{ij}$ 
is neither constant nor single valued.
The multiple-value ambiguity does not affect the 
physics of the low energy effective action since it may be related to 
electric-magnetic duality, as shown by Seiberg and Witten [1]. 

\medskip

A natural setting in which the above monodromy problem may be solved is 
provided by families of Riemann surfaces, called the Seiberg-Witten spectral
curves, 
denoted by $\Gamma$. Indeed, letting the quantum moduli $a_j$ correspond to 
moduli of the Riemann surfaces, there is automatically a complex analytic 
period matrix, whose imaginary part is positive definite, and whose monodromy 
group corresponds to the modular group of the surface.
The general set-up of the Seiberg-Witten solution, expected for arbitrary gauge 
algebra $\G$ with rank $n$ and general hypermultiplet representation is as 
follows.

\medskip

\item{(1)} The Seiberg-Witten curve is a family of Riemann surfaces $\Gamma 
(u_1, \cdots , u_n)$ dependent on $n$ auxiliary complex parameters $u_j$, which
are related to the quantum moduli $a_j$. The Seiberg-Witten curve will also 
depend upon the gauge coupling $g$ and $\theta$-angle and on the hypermultiplet
masses $m_k$.

\item{(2)} The Seiberg-Witten differential 1-form $d\lambda$ is meromorphic
on $\Gamma$, with residues which are linear in the hypermultiplet masses $m_k$. Since 
the hypermultiplet masses receive no quantum corrections as $a_j$ varies,
the derivatives $\partial (d\lambda)/\partial a_j$ are holomorphic 1-forms.

\item{(3)} The quantum moduli and the prepotential are given by
$$
a_j = {1\over 2\pi i}\oint _{A_j} d\lambda \qquad \qquad
a_{Dj} = {\partial \F \over \partial a_j} = {1\over 2\pi i}\oint _{B_j} d 
\lambda
\eqno (2.5)
$$
for a suitable set of cycles $A_j$ and $B_j$ on $\Gamma$.

\medskip
Shortly after the initial work of Seiberg and Witten, curves and 
differentials were proposed for a general classical
gauge group, with and without hypermultiplets
in the fundamental representation.  
Use was made of the $R$-charge assignments of the fields, the singularity 
structure of the degenerations of  the Seiberg-Witten 
curve and much educated guess work (see e.g. [11] for reviews).
The monodromies and instanton corrections in the corresponding prepotential
were determined in [31].
More recently, Seiberg-Witten curves for many other
theories have been found, based on integrable models [12-14],
M-Theory [15], and geometric engineering [16].
Monodromies and instanton corrections
have been determined in several important cases,
including for
the $SU(N)$ gauge
theory, with hypermultiplet in the symmetric or
anti-symmetric representation [17].
The prepotential has also been analyzed using methods from
Whitham theory [18, 30] and WDVV equations [19]. 

\bigskip
\bigbreak

\centerline{\bf III. TWISTED AND UNTWISTED CALOGERO-MOSER SYSTEMS}

\bigskip

\noindent
{\bf a) The $SU(N)$ Elliptic Calogero-Moser System}

The original elliptic Calogero-Moser
system is the system defined by the Hamiltonian
$$
H(x,p)={1\over 2}\sum_{i=1}^Np_i^2-{1\over 2}m^2
\sum_{i\not=j}\wp(x_i-x_j)
\eqno(3.1)
$$
Here $m$ is a mass parameter, and $\wp(z)$ is the Weierstrass $\wp$-function,
defined on a torus ${\bf C}/(2\omega_1{\bf Z}+2\omega_2{\bf Z})$. As usual, we
denote by $\tau=\omega_2/\omega_1$ the moduli of the torus, and set $q=e^{2\pi 
i\tau}$. The well-known trigonometric and rational limits with respective
potentials 
$$
\eqalign{
H_{\rm trig} & = \half \sum _{i=1} ^N p_i ^2 -{1\over 2}m^2\sum_{i\not=j}
{1\over 4\,{\rm sh}^2\,({x_i-x_j\over 2})}
\cr
H_{\rm rat} & = \half \sum _{i=1} ^N p_i ^2
-{1\over 2}m^2\sum_{i\not=j}{1\over (x_i-x_j)^2}
\cr}
$$ 
arise in the limits $\omega_1=-i\pi,\omega_2\rightarrow\infty$
and $\omega_1,\omega_2\rightarrow\infty$.
All these systems have been shown to be completely integrable
in the sense of Liouville, i.e. they all admit a complete set of integrals 
of motion which are in involution [20-22]. For a recent review of some 
applications of these 
models see [23].

\medskip

Our considerations require however a notion of integrability which
is in some sense more stringent, namely the existence of a Lax pair $L(z)$,
$M(z)$ with spectral parameter $z$. Such a Lax pair was obtained by Krichever
[24] in 1980. He showed that the Hamiltonian system (3.1) is
equivalent to the Lax equation $\dot L(z)=[L(z),M(z)]$,
with $L(z)$ and $M(z)$ given by the following $N\times N$ matrices
$$
\eqalignno{
L_{ij}(z)=&p_i\d_{ij}-m(1-\d_{ij})\P(x_i-x_j,z)\cr
M_{ij}(z)=&m\d_{ij}\sum_{k\not= i}\wp(x_i-x_k)-m(1-\d_{ij})\P'(x_i-x_j,z).
&(3.2)\cr}
$$ 
The function $\P(x,z)$ is defined by
$$
\P(x,z)={\sigma(z-x)\over
\sigma(z)\sigma(x)}e^{x\zeta(z)},
\eqno(3.3)
$$
where $\sigma(z)$, $\zeta(z)$ are the usual Weierstrass $\sigma$ and $\zeta$
functions on the torus ${\bf C}/(2\omega_1{\bf Z}+2\omega_2{\bf  Z})$.
The function $\P(x,z)$ satisfies the key functional equation
$$
\Phi(x,z)\P'(y,z)
-\P(y,z)\P'(x,z)
=(\wp(x)-\wp(y))\P(x+y,z).
\eqno(3.4)
$$ 
It is well-known that functional equations of this form 
are required for the Hamilton
equations of motion to be equivalent to the Lax equation
$\dot L(z)=[L(z),M(z)]$ with a Lax pair of the form (3.2).
It is a relatively recent result
of Braden and Buchstaber [25] that, conversely,
general functional equations of the form (3.4) essentially
determine $\P(x,z)$.

\bigskip

\noindent
{\bf b) Calogero-Moser Systems defined by Lie Algebras}

Olshanetsky and Perelomov [6] showed that
the Hamiltonian system (3.1) is only one
of a series associated with each simple Lie algebra.
Given any simple Lie algebra $\G$,
Olshanetsky and Perelomov [6] introduced the system
with Hamiltonian
$$
H(x,p)
={1\over 2}\sum_{i=1}^rp_i^2
-{1\over 2}
\sum_{\a\in{\cal R}(\G)}
m_{|\a|}^2\wp(\a\cdot x),
\eqno(3.5)
$$
where $r$ is the rank of $\G$, ${\cal R}(\G)$
denotes the set of roots of $\G$, and the $m_{|\a|}$ are mass parameters.
To preserve the invariance of (3.5)
under the Weyl group, the parameters $m_{|\a|}$ depend only
on the orbit $|\a|$ of the root $\a$, and not on the root $\a$ itself.
In the case of $A_{N-1}= SU(N)$, it is common practice to use $N$ pairs of
dynamical variables $(x_i,p_i)$, since the roots of $A_{N-1}$
lie conveniently on a hyperplane in ${\bf C}^N$.
The dynamics of the system are unaffected if we shift
all $x_i$ by a constant, and the number of degrees of freedom
is effectively $N-1=r$. Now the roots of $SU(N)$ are given
by $\a=e_i-e_j$, $1\leq i,j\leq N$, $i\not=j$. Thus
we recognize the original elliptic Calogero-Moser system
as the special case of (3.5) corresponding to $A_{N-1}$.
Olshanetsky and Perelomov constructed a
Lax pair for all these systems with {\it classical}
Lie algebras, without spectral parameter [6].

\bigskip

\noindent
{\bf c) Twisted Calogero-Moser Systems defined by Lie Algebras}

It turns out that the Hamiltonian systems (3.5) are not the
only natural extensions of the basic elliptic Calogero-Moser
system. 
A subtlety arises for simple Lie algebras $\G$ which are not
simply-laced, i.e., algebras which admit roots of uneven
length. This is the case for the algebras $B_n$, $C_n$, $G_2$,
and $F_4$ in Cartan's classification.
For these algebras, the following {\it twisted} elliptic
Calogero-Moser systems were introduced by the authors in [7,8]
$$
H^{{\rm twisted}}
=
{1\over 2}\sum_{i=1}^rp_i^2
-{1\over 2}
\sum_{\a\in{\cal R}(\G)}
m_{|\a|}^2
\wp_{\nu(\a)}(\a\cdot x).
\eqno(3.6)
$$
Here the function $\nu(\a)$ depends only on the length of the root $\a$.
If $\G$ is simply-laced, we set $\nu(\a)=1$ identically. Otherwise,
for $\G$ non simply-laced, we set $\nu(\a)=1$ when $\a$ is a long root,
$\nu(\a)=2$ when $\a$ is a short root and $\G$ is one of the
algebras $B_n$, $C_n$, or $F_4$, and $\nu(\a)=3$ when $\a$ is a short root
and $\G=G_2$. The {\it twisted} Weierstrass function $\wp_{\nu}(z)$
is defined by
$$
\wp_{\nu}
(z)
=\sum_{\sigma=0}^{\nu-1}
\wp(z+2\omega_a{\sigma\over\nu}),
\eqno(3.7)
$$
where $\omega_a$ is any of the half-periods $\omega_1$,
$\omega_2$, or $\omega_1+\omega_2$.
Thus the twisted and untwisted Calogero-Moser systems coincide
for $\G$ simply laced. 
The original motivation for twisted
Calogero-Moser systems was based on
their scaling limits
(which will be discussed in the next section) [7,8].
Another motivation based on the symmetries of Dynkin diagrams
was proposed subsequently by Bordner, Sasaki, and Takasaki [26].

\bigskip
\bigbreak

\centerline{\bf IV. SCALING LIMITS OF CALOGERO-MOSER SYSTEMS}

\bigskip

\noindent
{\bf a) Results of Inozemtsev for $A_{N-1}$}

For the standard elliptic Calogero-Moser systems
corresponding to $A_{N-1}$, Inozemtsev [27] has shown in the
1980's that in the scaling limit
$$
\eqalign{
m   & = Mq^{-{1\over 2N}},\ \ \ \qquad  q\rightarrow 0 \cr
x_i & = X_i-2\omega_2{i\over N},
\qquad
1\leq i\leq N \cr}
\eqno (4.1)
$$
where $M$ is kept fixed, the elliptic $A_{N-1}$ Calogero-Moser Hamiltonian 
tends to the following Hamiltonian 
$$
H_{{\rm Toda}}
={1\over 2}\sum_{i=1}^Np_i^2
-{1\over 2}\big(\sum_{i=1}^{N-1}e^{X_{i+1}-X_i}+e^{X_1-X_N}\big)
\eqno(4.2)
$$
The roots $e_i-e_{i+1}$, $1\leq i\leq N-1$, and $e_N-e_1$
can be recognized as the simple roots of the
affine algebra $A_{N-1}^{(1)}$.
(For basic facts on affine algebras, we refer to [28]).
Thus (4.2) can be recognized as the Hamiltonian of the Toda system
defined by $A_{N-1}^{(1)}$.

\bigskip

\noindent
{\bf b) Scaling Limits based on the Coxeter Number}

The key feature of the above scaling limit is the collapse
of the sum over the entire root lattice of $A_{N-1}$
in the Calogero-Moser Hamiltonian to the
sum over only simple roots in the Toda Hamiltonian for the
Kac-Moody algebra $A_{N-1}^{(1)}$.
Our task is to extend this mechanism to general Lie algebras.
For this, we consider the following generalization  of the preceding scaling
limit
$$
\eqalign{
m & = Mq^{-{1\over 2}\d},\cr
x & = X-2\omega_2\d\rho^{\vee},\cr}
\eqno (4.3)
$$
Here $x=(x_i)$, $X=(X_i)$ and $\rho^{\vee}$
are $r$-dimensional vectors.
The vector $x$ is the dynamical
variable of the Calogero-Moser system.
The parameters $\d$ and $\rho^{\vee}$
depend on the algebra $\G$ and are yet to be chosen.
As for $M$ and $X$, they have the same interpretation as
earlier, namely as respectively the mass parameter
and the dynamical variables of the limiting system.
Setting $\omega_1=-i\pi$,
the contribution of each root $\a$ to the Calogero-Moser
potential can be expressed as
$$
m^2\wp(\a\cdot x)
=
{1\over 2}M^2
\sum_{n=-\infty}^{\infty}
{e^{2\d\omega_2}\over
{\rm ch}(\a\cdot x-2n\omega_2)-1}
\eqno(4.4)
$$
It suffices to consider positive roots $\a$.
We shall also assume that $0\leq \d\,\a\cdot\rho^{\vee}
\leq 1$. The contributions of the $n=0$ and $n=-1$
summands in (4.4) are proportional
to $e^{2\omega_2(\d-\d\,\a\cdot\rho^{\vee})}$
and $e^{2\omega_2(\d-1+\d\,\a\cdot\rho^{\vee})}$
respectively.
Thus the existence of a finite scaling limit requires
that $\d\,\leq\d\,\a\cdot\rho^{\vee}\leq 1-\d.$
Let $\a_i$, $1\leq i\leq r$ be a basis of simple roots
for $\G$. If we want all simple roots $\a_i$
to survive in the limit, we must require that
$\a_i \cdot\rho^{\vee}=1,\ 1\leq i\leq r.$
This condition characterizes the vector $\rho^{\vee}$
as the {\it level vector}.
Next, the second condition in (3.7)
can be rewritten as $\d\{1+max_{\a}\,(\a\cdot\rho^{\vee})\}
\leq 1$. But
$$
h_{\G}=1+max_{\a}\,(\a\cdot\rho^{\vee})
\eqno(4.5)
$$
is precisely the Coxeter number of $\G$,
and we must have $\d\leq {1\over h_{\G}}$.
Thus when $\d<{1\over h_{\G}}$,
the contributions of all the roots except
for the simple roots of $\G$ tend to $0$.
On the other hand, when $\d={1\over h_{\G}}$,
the highest root $\a_0$ realizing the maximum over
$\a$ in (4.5) survives.
Since $-\a_0$ is the additional
simple root for the affine Lie algebra
$\G^{(1)}$, we arrive in this way at the following theorem,
which was proved in [8]

\bigskip
\noindent
{\bf Theorem 1}.
{\it Under the limit (4.4-4.5), with $\d={1\over h_{\G}}$,
and $\rho^{\vee}$ given by the
level vector,
the Hamiltonian of the elliptic Calogero-Moser system
for the simple Lie algebra $\G$
tends to the Hamiltonian of the Toda system
for the affine Lie algebra $\G^{(1)}$.}

\bigskip

\noindent
{\bf (c) Scaling Limit based on the Dual Coxeter Number}

If the Seiberg-Witten spectral curve of the $\N=2$
supersymmetric gauge theory with a hypermultiplet in
the adjoint representation is to be realized as
the spectral curve for a Calogero-Moser system,
the parameter $m$ in the Calogero-Moser system
should correspond to the mass of the hypermultiplet.
In the gauge theory, the dependence of the coupling
constant on the mass $m$ is given by
$$
\tau={i\over 2\pi}h_{\G}^{\vee}{\rm ln}\,{m^2\over M^2}
\qquad
\Longleftrightarrow
\qquad
m=Mq^{-{1\over 2h_{\G}^{\vee}}}
\eqno(4.6)
$$
where $h_{\G}^{\vee}$ is the quadratic Casimir of the
Lie algebra $\G$. This shows that the correct physical
limit, expressing the decoupling of the hypermultiplet
as it becomes infinitely massive,
is given by (4.3), but with $\d={1\over h_{\G}^{\vee}}$.
To establish a closer parallel with our preceding discussion,
we recall that the quadratic Casimir $h_{\G}^{\vee}$
coincides with the {\it dual Coxeter number} of $\G$,
defined by
$$
h_{\G}^{\vee}=1+max_{\a}\,(\a^{\vee}\cdot\rho),
\eqno(4.7)
$$
where $\a^{\vee}={2\a\over\a^2}$ is the coroot associated
to $\a$, and $\rho={1\over 2}\sum_{\a>0}\a$
is the Weyl vector.

\medskip

For simply-laced Lie algebras $\G$ (ADE algebras),
we have $h_{\G}=h_{\G}^{\vee}$, and the preceding scaling limits
apply. However, for non simply-laced algebras
($B_n$, $C_n$, $G_2$, $F_4$), 
we have $h_{\G}>h_{\G}^{\vee}$,
and our earlier considerations show that the untwisted
elliptic Calogero-Moser Hamiltonians do not tend to
a finite limit under (4.6), $q\to 0$,
$M$ is kept fixed.
This is why the twisted Hamiltonian systems (3.6)
have to be introduced. The twisting produces precisely
an improvement in the asymptotic behavior
of the potential which allows a finite, non-trivial limit.
More precisely,
we can write
$$
m^2\wp_{\nu}(x)
=
{\nu^2\over 2}
\sum_{n=-\infty}^{\infty}
{m^2\over {\rm ch}\,\nu(x-2n\omega_2)-1}.
\eqno(4.8)
$$
Setting $x=X-2\omega_2\d^{\vee}\rho$, we
obtain the following asymptotics
$$
m^2\wp_{\nu}(x)
=\nu^2M^2
\cases{e^{-2\omega_2(\d^{\vee}\a^{\vee}\cdot\rho-\d^{\vee})-\a^{\vee}\cdot 
X}
+e^{-2\omega_2(1-\d^{\vee}\a^{\vee}\cdot\rho-\d^{\vee})+\a^{\vee}\cdot X},
&if $\a$ is long;\cr
e^{-2\omega_2(\d^{\vee}\a^{\vee}\cdot\rho-\d^{\vee})-\a^{\vee}\cdot X},
&if $\a$ is short.\cr}
$$
This leads to the following theorem [8]

\bigskip
\noindent
{\bf Theorem 2}.
{\it 
Under the limit $x=X+2\omega_2{1\over h_{\G}^{\vee}}\rho$,
$m=Mq^{-{1\over 2h_{\G}^{\vee}}}$,
with $\rho$ the Weyl vector and $q\to 0$,
the Hamiltonian of the twisted elliptic Calogero-Moser system
for the simple Lie algebra $\G$
tends to the Hamiltonian of the Toda system
for the affine Lie algebra $(\G^{(1)})^{\vee}$.}

\bigskip

Similar arguments show that the Lax pairs constructed below
also have finite, non-trivial scaling limits whenever
this is the case for the Hamiltonians.

\bigskip

\centerline{\bf V. LAX PAIRS FOR CALOGERO-MOSER SYSTEMS}

\bigskip

\noindent
{\bf a) The General Ansatz}

Let the rank of $\G$ be $n$,
and $d$ be its dimension.
Let $\L$ be a representation of $\G$ of dimension $N$,
of weights $\l_I$, $1\leq I\leq N$. Let $u_I\in {\bf C}^N$
be the weights of the fundamental
representation of $GL(N,{\bf C})$. Project
orthogonally the $u_I$'s onto the $\l_I$'s as
$$
su_I=\l_I+u_I,
\ \qquad
\l_I\perp v_J.
\eqno(5.1)
$$
It is easily verified that $s^2$ is the second Dynkin index.
Then $\a_{IJ}=\l_I-\l_J$
is a weight of $\L\otimes\L^*$ associated to the
root $u_I-u_J$ of $GL(N,{\bf C})$. The Lax pairs
for both untwisted and twisted Calogero-Moser systems
will be of the form
$$
L=P+X,
\ \
M=D+X,
\eqno(5.2)
$$
where the matrices $P,X,D$, and $Y$ are given by
$$
X=\sum_{I\not=J}C_{IJ}\P_{IJ}(\a_{IJ},z)E_{IJ},
\ \ \
Y=\sum_{I\not=j}C_{IJ}\P'_{IJ}(\a_{IJ},z)E_{IJ}
\eqno(5.3a)
$$
and by
$$
P=p\cdot h,
\ \ \ \
D=d\cdot (h\oplus\tilde h)+\Delta.
\eqno(5.3b)
$$
Here $h$ is in a Cartan subalgebra ${\cal H}_{\G}$
for $\G$, $\tilde h$
is in the Cartan-Killing orthogonal
complement of ${\cal H}_{\G}$
inside a Cartan subalgebra ${\cal H}$ for $GL(N,{\bf C})$,
and $\Delta$ is in the centralizer of ${\cal H}_{\G}$
in $GL(N,{\bf C})$.
The functions $\P_{IJ}(x,z)$ and the coefficients
$C_{IJ}$ are yet to be determined.
We begin by stating the necessary and sufficient
conditions for the pair $L(z)$, $M(z)$ of (5.2)
to be a Lax pair for the
(twisted or untwisted) Calogero-Moser
systems. For this, it is convenient to
introduce the following notation
$$
\eqalignno{
\P_{IJ}&=\P_{IJ}(\a_{IJ}\cdot x)\cr
\wp_{IJ}'
&=\P_{IJ}(\a_{IJ}\cdot x,z)\P_{JI}'(-\a_{IJ}\cdot x,z)
-\P_{IJ}(-\a_{IJ}\cdot x,z)
\P_{JI}'(\a_{IJ}\cdot x,z).
&(5.4)
\cr}
$$

Then the Lax equation $\dot L(z)
=[L(z),M(z)]$ implies the
Calogero-Moser system if and only
if the following three identities are satisfied
$$
\sum_{I\not=J}C_{IJ}C_{JI}\wp_{IJ}'\a_{IJ}
=
s^2\sum_{\a\in {\cal R}(\G)}
m_{|\a|}^2\wp_{\nu(\a)}(\a\cdot x)
\eqno(5.5a)
$$
$$
\sum_{I\not=J}C_{IJ}C_{JI}
\wp_{IJ}'(v_I-v_J)
=0
\eqno(5.5b)
$$
$$
\eqalignno{
\sum_{K\not= I,J}
C_{IK}C_{KJ}(\P_{IK}\P_{KJ}'-\P_{IK}'\P_{KJ})
&=
sC_{IJ}\P_{IJ}d\cdot (v_I-v_J)
+
\sum_{K\not= I,J}
\Delta_{IJ}C_{KJ}\P_{KJ}\cr
&\qquad\qquad\qquad
-
\sum_{K\not= I,J}
C_{IK}\P_{IK}\Delta_{KJ}
&(5.5c)\cr}
$$
\noindent
The following theorem was established in [7]:

\bigskip

\noindent
{\bf Theorem 3}. {\it A representation $\Lambda$, functions
$\Phi_{IJ}$, and coefficients $C_{IJ}$ with a spectral parameter $z$
satisfying (5.5a-c) can be found for all twisted and untwisted elliptic
Calogero-Moser systems associated with a simple Lie algebra
$\G$, except possibly in the case of twisted $G_2$.
In the case of $E_8$, we have to assume the existence of 
a $\pm1$ cocycle.}   

\bigskip

\noindent
{\bf b) Lax Pairs for Untwisted Calogero-Moser Systems}

We now describe some important
features of the Lax pairs we obtain in this manner. All 
have an independent spectral parameter.

\bigskip

$\bullet$ In the case of the {\it untwisted} Calogero-Moser systems,
we can choose $\P_{IJ}(x,z)=\P(x,z)$, $\wp_{IJ}(x)=\wp(x)$ for all $\G$. One 
also has $\Delta=0$ for all $\G$, except for $E_8$.

\medskip

$\bullet$ For $A_n$, the Lax pair (5.2-5.3) may correspond
to any of the totally antisymmetric representations, including the
fundamental one.
 
\medskip

$\bullet$ For the $BC_n$ system, the Lax pair is
obtained by imbedding $B_n$ in $GL(N,{\bf C})$
with $N=2n+1$. When $z=\omega_a$ (half-period),
the Lax pair obtained this way reduces to
that of Olshanetsky and Perelomov [6].

\medskip 

$\bullet$ For the $B_n$ and $D_n$ systems,
additional Lax pairs are found by taking $\L$ to be
the spinor representation.

\medskip

$\bullet$ For $G_2$,
a Lax pair is obtained in the representation ${\bf 7}$ 
while another one is gotten by restricting the {\bf 8}
of $B_3$ to the ${\bf 7}\oplus{\bf 1}$ of $G_2$.

\medskip

$\bullet$ For $F_4$, a Lax pair can be obtained by
taking $\L$ to be the ${\bf 26}\oplus{\bf 1}$
of $F_4$, viewed as the restriction of
the {\bf 27} of $E_6$ to its $F_4$ subalgebra.

\medskip

$\bullet$ For $E_6$, $\L$ is the {\bf 27} representation.

\medskip

$\bullet$ For $E_7$, $\L$ is the {\bf 56} representation.

\medskip

$\bullet$ For $E_8$, a Lax pair with spectral parameter
can be constructed with $\L$ given by the {\bf 248} representation,
if coefficients $c_{IJ}=\pm 1$ exist
with the following cocycle conditions
$$
\eqalignno{
c(\lambda,\lambda-\d)c(\lambda-\d,\mu)=&
c(\lambda,\mu+\d)c(\mu+\d,\mu)\cr
&{\rm \ when\ \d\cdot\lambda=-\d\cdot\mu=1,
\
\lambda\cdot\mu=0}\cr
c(\l,\mu)c(\l-\d,\mu)=&c(\l,\l-\d)\cr
& {\rm \ when\ \d\cdot\l=\l\cdot\mu=1,
\
\d\cdot\mu=0}\cr
c(\l,\mu)c(\l,\l-\mu)=&
-c(\l-\mu,-\mu)\cr
&{\rm \ when\ \l\cdot\mu=1}.
&(5.6a)
\cr}
$$
The matrix $\Delta$ in the Lax pair is then the $8\times 8$ matrix
given by
$$
\eqalignno{
\Delta_{ab}=&
\sum_{\d\cdot\b_a=1\atop \d\cdot\b_b=1}
{m_2\over 2}
\big(c(\b_a,\d)c(\d,\b_b)
+
c(\b_a,\b_a-\d)c(\b_a-\d,\b_b)\big)
\wp(\d\cdot x)\cr
&
-\sum_{\d\cdot\b_a=1\atop \d\cdot\b_b=-1}
{m_2\over 2}
\big(c(\b_a,\d)c(\d,\b_b)
+
c(\b_a,\b_a-\d)c(\b_a-\d,\b_b)\big)
\wp(\d\cdot x)\cr
\Delta_{aa}=&
\sum_{\b_a\cdot\d=1}
m_2\wp(\d\cdot x)
+2m_2\wp(\b_a\cdot x),
&(5.6b)
\cr}
$$
where $\b_a$, $1\leq a\leq 8$, is a maximal set of 8 mutually
orthogonal roots.

\medskip

$\bullet$ Explicit expressions for the constants $C_{IJ}$ and the functions 
$d(x)$,
and thus for the Lax pair are particularly simple when the representation
$\Lambda$ consists of only a single Weyl orbit of weights. This is the case 
when
$\Lambda$ is either

\medskip

\item{(1)} the defining representation of $A_n$, $C_n$ or $D_n$;

\item{(2)} any rank $p$ totally anti-symmetric representation of $A_n$;

\item{(3)} an irreducible fundamental spinor representation of $B_n$ or $D_n$;

\item{(4)} the ${\bf 27}$ of $E_6$; 

\item{(5)} the ${\bf 56}$ of $E_7$.

\medskip
Then the weights $\lambda$ and $\mu$ of $\Lambda$ provide unique labels instead
of $I$ and $J$, and the values of $C_{IJ}=C_{\lambda \mu}$ are given by a
simple formula
$$
C_{\lambda \mu} = \left \{ \matrix{
\sqrt{{\alpha ^2 \over 2}} m_{|\alpha|} &
{\rm when} \ \alpha =\lambda - \mu \ {\rm is \ a \ root} \cr
&\cr
0 & {\rm otherwise} \cr} \right .
\eqno (5.7)
$$

The expression for the vector $d$ may be summarized by
$$
s d \cdot u_\lambda = \sum _{\lambda \cdot \delta =1;\ \delta ^2=2}
m_{|\delta |} \wp (\delta \cdot x)
\eqno (5.8)
$$
(For $C_n$, the last equation has an additional term, as given in [7].)
In each case, the number of independent couplings $m_{|\alpha|}$ equals the
number of different root lengths.

\bigskip

\noindent
{\bf c) Lax Pairs for Twisted Calogero-Moser Systems}

Recall that the twisted and untwisted Calogero-Moser systems
differ only for non-simply laced Lie algebras, namely
$B_n$, $C_n$, $G_2$ and $F_4$.
These are the only algebras we discuss in
this paragraph.
The construction (5.2-5.5) gives then
Lax pairs for all of them, 
with the possible exception of twisted $G_2$.
Unlike the case of untwisted Lie algebras however,
the functions $\P_{IJ}$ have to be chosen
with care, and differ for each algebra.
More specifically,

\medskip

$\bullet$ For $B_n$, the Lax pair is of dimension $N=2n$,
admits two independent couplings $m_1$ and $m_2$,
and
$$
\P_{IJ}(x,z)
=
\cases{
\P(x,z), &if $I-J\not= 0,\pm n$\cr
\P_2({1\over 2}x,z), &if $I-J=\pm n$\cr}.
\eqno(5.9)
$$
Here a new function $\P_2(x,z)$ is defined by
$$
\P_2({1\over 2}x,z)
={\P({1\over 2}x,z)\P({1\over 2}x+\omega_1,z)
\over
\P(\omega_1,z)}
\eqno(5.10)
$$

$\bullet$ For $C_n$, the Lax pair is of dimension $N=2n+2$,
admits one independent coupling $m_2$,
and
$$
\P_{IJ}(x,z)
=
\P_2(x+\omega_{IJ},z),
$$
where $\omega_{IJ}$ are given by
$$
\omega_{IJ}
=
\cases{0, &if $I\not=J=1,2,\cdots,2n+1$;\cr
\omega_2, &if $1\leq I\leq 2n,\ J=2n+2$;\cr
-\omega_2, &if $1\leq J\leq 2n,\ I=2n+2$.\cr}
\eqno(5.11)
$$

$\bullet$ For $F_4$, the Lax pair is of dimension $N=24$,
admits two independent couplings $m_1$ and $m_2$,
$$
\P_{\l\mu}(x,z)
=
\cases{\P(x,z), &if $\l\cdot\mu=0$;\cr
\P_1(x,z), &if $\l\cdot\mu={1\over 2}$;\cr
\P_2({1\over 2}x,z), &if $\l\cdot\mu=-1$.\cr}
\eqno(5.12)
$$
where the function $\P_1(x,z)$ is defined by
$$
\P_1(x,z)
=
\P(x,z)
-
e^{\pi i\zeta(z)+\eta_1z}
\P(x+\omega_1,z)
\eqno(5.13)
$$
Here it is more convenient to label
the entries of the Lax pair directly by the weights
$\lambda=\lambda_I$ and
$\mu=\lambda_J$ instead of $I$ and $J$.

$\bullet$ For $G_2$, there are natural candidates for Lax pairs
in the {\bf 6} and {\bf 8}
representations of $G_2$, but it is still unknown whether
elliptic functions $\P_{IJ}(x,z)$
exist which satisfy the required identities.

\medskip

We note that recently Lax pairs of root type have been considered [21] which
correspond, in the above Ansatz (5.3-5), to $\Lambda$ equal to the adjoint
representation of $\G$ and the coefficients $C_{IJ}$ vanishing for $I$ or $J$
associated with zero weights. This choice yields another Lax pair for the case
of $E_8$.

\bigskip
\bigbreak

\centerline{\bf VI. CALOGERO-MOSER AND SEIBERG-WITTEN THEORY}

\bigskip

The correspondence between Seiberg-Witten 
theory for $\N=2$ super-Yang-Mills theory with one hypermultiplet in the 
adjoint representation of the gauge algebra, and the elliptic Calogero-Moser 
systems was first established in [5], for the gauge algebra $\G=SU(N)$.
We describe it here in some detail.

\bigskip

\noindent
{\bf a) The Case of $\G = SU(N)$}

All that we shall need here of the elliptic Calogero-Moser system is its Lax
operator $L(z)$, whose $N\times N$ matrix elements are given by
$$
L_{ij}(z) = p_i\delta _{ij} - m(1-\delta _{ij}) \Phi (x_i-x_j,z)
\eqno (6.1)
$$
Notice that the Hamiltonian is simply given in terms of $L$ by $H(x,p) =
\half \tr L(z)^2 + C\wp (z)$ with $C=-\half m^2 N(N-1)$.
The correspondence between the data of the elliptic Calogero-Moser system and 
those of the Seiberg-Witten theory is as follows. 

\item{(1)} The parameter $m$ in (6.1) is the hypermultiplet mass;

\item{(2)} The gauge coupling $g$ and the $\theta$-angle are related to the 
modulus of the torus $\Sigma ={\bf C} /(2\omega _1 {\bf Z} + 2 \omega _2 {\bf 
Z})$
by 
$$
\tau = {\omega _2 \over \omega _1} = {\theta \over 2 \pi} + {4 \pi i\over g^2}
\, ;
\eqno (6.2)
$$

\item{(3)} The Seiberg-Witten curve $\Gamma$ is the spectral curve of the 
elliptic Calogero-Moser model, 
$$
\Gamma = \{ (k,z) \in {\bf C} \times \Sigma, \ \det\bigl (kI-L(z)\bigr )=0\}
\eqno (6.3)
$$
The Seiberg-Witten form is $d\lambda = k \ dz$. $\Gamma$ is invariant 
under the Weyl group of $SU(N)$. 

\item{(4)} Using the Lax equation $\dot L = [L,M]$, it is clear that the 
spectral curve is independent of time, and can be dependent only upon the 
constants of motion of the Calogero-Moser system, of which there are only $N$.
These integrals of motion may be viewed as parametrized by the quantum moduli
of the Seiberg-Witten system.

\item{(5)} Finally, $d\lambda =kdz$ is meromorphic, with a simple pole on 
each of the $N$ sheets above the point $z=0$ on the base torus. The residue at 
each of these poles is proportional to $m$, as required by the general set-up 
of Seiberg-Witten theory, explained in \S II.

\bigskip

\noindent
{\bf b) Four Fundamental Results for the Case of $\G = SU(N)$}

While the above mappings of the Seiberg-Witten data onto the Calogero-Moser 
data is certainly natural, there is no direct proof of it, and it is important 
to check that the results inferred from it agree with known facts from quantum 
field theory. To establish this, as well as a series of further predictions
from the correspondence, we give four theorems (the
proofs may be found in [5] for the first three theorems,
and in [29] for the last one).

\bigskip

\noindent
{\bf Theorem 4.} {\it The spectral curve equation $\det (kI-L(z))=0$ is 
equivalent to
$$
\vartheta _1 \biggl (
{1 \over 2\omega _1}(z-m{\partial \over \partial k}) \big | \tau \biggr ) 
H(k)=0
\eqno (6.4)
$$
where $H(k)$ is a monic polynomial in $k$ of degree $N$, whose zeros (or 
equivalently whose coefficients) correspond to the moduli of the gauge theory. 
If $H(k)=\prod_{i=1}^N(k-k_i)$, then
$$
\lim_{q\to 0}{1\over 2\pi i}\oint_{A_i}kdz=k_i-{1\over 2}m.
$$}
Here, $\vartheta _1$ is the Jacobi $\vartheta$-function, which admits a simple 
series expansion in powers of the instanton factor $q=e^{2\pi i \tau}$, so that 
the curve equation may also be rewritten as a series expansion
$$
\sum _{n \in {\bf Z}} (-)^n q ^{\half n(n-1)} e^{nz} H(k-n \cdot m) =0
\eqno (6.5)
$$
where we have set $\omega _1 = -i\pi$ without loss of generality. 

\bigskip

\noindent
{\bf Theorem 5.} {\it The prepotential of the Seiberg-Witten theory obeys a 
renormalization group-type equation that simply relates $\F$ to the 
Calogero-Moser Hamiltonian, expressed in terms of the quantum order parameters 
$a_j$
$$
a_j = {1\over 2\pi i}\oint _{A_j} d \lambda
\qquad \qquad
{\partial \F \over \partial \tau} \bigg | _{a_j } 
= H(x,p) = \half \tr L(z)^2 +C \wp (z)
\eqno (6.6)
$$
Furthermore, in an expansion in powers of the instanton factor $q=e^{2\pi i 
\tau}$, the quantum order parameters $a_j$ may be computed by residue methods
in terms of the zeros of $H(k)$.}

\bigskip

The proof of (6.6) requires Riemann surface deformation
theory [5]. The fact that the quantum order parameters may 
be evaluated by residue methods arises from the fact that $A_j$-cycles may be 
chosen on the spectral curve $\Gamma$ in such a way that they will shrink to 
zero as $q \to 0$. As a result, contour integrals around full-fledged branch 
cuts $A_j$ reduce to contour integrals around poles at single points, which may 
be calculated by residue methods only. These methods were originally developed
in [30,31]. Knowing the quantum order parameters in  terms of the zeros $k_j$ 
of
$H(k)=0$ is a relation which may be inverted and  used in (6.6) to obtain a
differential relation for all order instanton  corrections. It is now only
necessary to evaluate explicitly the 
$\tau$-independent contribution to $\F$, which in field theory arises from 
perturbation theory. This may be done easily by retaining only the $n=0$ and 
$n=1$ terms in the expansion of the curve (6.5), so that 
$ z= \ln H(k) - \ln H(k-m)$. The results of the calculations to two instanton 
order may be summarized in the following theorem [5].

\bigskip

\noindent
{\bf Theorem 6.} {\it The prepotential, to 2 instanton order is given by $\F = 
\F ^{({\rm pert})} + \F ^{(1)} + \F ^{(2)}$. The perturbative contribution is 
given by
$$
\F ^{({\rm pert})} = {\tau \over 2} \sum _i a_i ^2
 - {1 \over 8 \pi i} \sum _{i,j} \biggl [ (a_i -a_j)^2 \ln (a_i - a_j)^2
 - (a_i -a_j -m)^2 \ln (a_i -a_j -m)^2 \biggr ]
 \eqno (6.7a)
$$
while all instanton corrections are expressed in terms of a single 
function
$$
S_i(a) = {\prod _{j=1} ^N \big [ (a_i-a_j)^2-m^2 \bigr ] \over 
\prod _{j\not=i} (a-a_j)^2}
\eqno (6.7b)
$$
as follows}
$$
\eqalign{
\F ^{(1)} & = {q \over 2 \pi i} \sum _i S_i(a_i) \cr
\F ^{(2)} & = {q^2 \over 8 \pi i} \biggl [
  \sum _i S_i(a_i) \partial _i ^2 S_i(a_i)
  + 4 \sum _{i\not= j} {S_i(a_i)S_j(a_j) \over (a_i -a_j)^2} - { 
S_i(a_i)S_j(a_j) \over (a_i -a_j-m)^2}  \biggr ] \cr}
  \eqno (6.7c)
$$
The perturbative corrections to the prepotential of (6.7a) indeed precisely 
agree with the predictions of asymptotic freedom. The formulas
(6.7c) for the instanton corrections 
$\F^{(1)}$ and $\F^{(2)}$ are new, as they
have not yet been computed by direct field theory methods.
Perturbative expansions of the prepotential in powers
of $m$ have also been obtained in [32]. Within the context of 
topological gauge theory, these results on the prepotential have been 
used in [33].

\bigskip

The moduli $k_i$, $1\leq i\leq N$, of the gauge theory are 
evidently integrals of motion
of the system. To identify these integrals of motion, denote by $S$ 
any subset of $\{1,\cdots,N\}$, and let $S^*=\{1,\cdots,N\}\setminus S$,
$\wp(S)=\wp(x_i-x_j)$ when $S=\{i,j\}$. Let also $p_S$ denote
the subset of momenta $p_i$ with $i\in S$,
and $\sigma_k(p_S)$ the $k$-th symmetric polynomial
in $p_i$, $i\in S$. We have [29]

\bigskip

\noindent
{\bf Theorem 7}. {\it For any $K$, $0\leq K\leq N$,
let $\sigma_K(k_1,\cdots,k_N)=
\sigma_K(k)$ be the $K$-th symmetric polynomial of $(k_1,\cdots,k_N)$,
defined by $H(u)=\sum_{K=0}^N(-)^K\sigma_K(k)u^{N-K}$. Then}
$$
\sigma_K(k)
=
\sigma_K(p)
+
\sum_{l=1}^{[K/2]}m^{2l}
\sum_{|S_i\cap S_j|=2\delta_{ij}\atop 1\leq i,j\leq l}
\sigma_{K-2l}(p_{(\cup_{i=1}^lS_i)^*})
\prod_{i=1}^l[\wp(S_i)+{\eta_1\over\omega_1}]
\eqno(6.8)
$$

We note that an alternative derivation of (6.4) was recently presented in [34].
The parametrization (6.4) has also been extended to the spectral curves
of spin Calogero-Moser systems in [10]. For
spin Calogero-Moser systems with $l$ internal
degrees of freedom, we can write 
$$
det(\l I-L(z))=
\sum_{p=1}^l\p_z^{p-1}\big({\theta_1({1\over 2\omega_1}(z-m{\p\over\p k})|\tau)
\over
\theta_1({z\over 2\omega_1}|\tau)}H_p(k)\big)_{\vert_{k=\l+m\p_z{\rm log}\theta_1
({z\over2\omega_1}|\tau)}},\eqno(6.9)
$$
where $H_p(k)$ is a polynomial of degree $N-p+1$, $1\leq p\leq l$,
$H_1(k)$ is monic, and $H_p(k)$ has no term of order $k^0$. 

\bigskip

\noindent
{\bf c) The Case of General Gauge Algebra $\G$}

\bigskip

We consider now the $\N=2$ supersymmetric gauge theory for
a general simple gauge algebra $\G$ and 
a hypermultiplet of mass $m$ in the adjoint representation. Then [9]

\medskip

$\bullet$ the Seiberg-Witten curve of the theory
is given by the spectral curve $\Gamma=\{(k,z)\in{\bf C}\times\Sigma;
\det(kI-L(z))=0\}$ of the {\it twisted} elliptic
Calogero-Moser system associated to the Lie algebra $\G$.
The Seiberg-Witten differential $d\lambda$ is given by $d\lambda=kdz$.

\medskip
$\bullet$ The function $R(k,z)=\det(kI-L(z))$ is polynomial
in $k$ and meromorphic in $z$. The spectral curve $\Gamma$
is invariant under the Weyl group of $\G$. It depends
on $n$ complex moduli, which can be thought
of as independent integrals of motion of the Calogero-Moser system.

\medskip
$\bullet$
The differential $d \lambda=kdz$ is meromorphic on $\Gamma$,
with simple poles. The position and residues of the poles are
independent of the moduli. The residues are linear
in the hypermultiplet mass $m$. (Unlike the case of
$SU(N)$, their exact values
are difficult to determine for general $\G$).

\medskip
$\bullet$ In 
the $m \to 0$ limit, the Calogero-Moser system reduces to
a free system, the spectral curve $\Gamma$ is just the product
of several unglued copies of the base torus $\Sigma$,
indexed by the constant eigenvalues of $L(z)=p\cdot h$.
Let $k_i$, $1\leq i\leq n$, be
$n$ independent eigenvalues, and $A_i,B_i$ be the $A$
and $B$ cycles lifted to the corresponding sheets.
For each $i$, we readily obtain 
$$
\eqalign{
a_i & ={1\over 2\pi i}\oint_{A_i}d\lambda
={k_i\over 2\pi i}\oint_Adz={2\omega_1\over 2\pi i}k_i
\cr
a_{Di} & ={1\over 2\pi i}\oint_{B_i}d\lambda
={k_i\over 2\pi i}\oint_Bdz={2\omega_1\over 2\pi i}\tau k_i
\cr}
\eqno (6.9)
$$
Thus the prepotential $\F$ is given by
$\F={\tau\over 2}\sum_{i=1}^na_i^2$. This is the classical prepotential
and hence the correct
answer, since in the $m\to 0$ limit, the theory
acquires an $\N=4$ supersymmetry, and receives no
quantum corrections.

\medskip
$\bullet$ The $m\to\infty$ limit is the crucial consistency check,
which motivated the introduction of the {\it twisted} Calogero-Moser
systems in the first place [7,8]. In view of Theorem 2
and subsequent comments, in the limit $m\to\infty$, $q\to 0$,
with $x=X+2\omega_2{1\over h_{\G}^{\vee}}\rho$,
$m=Mq^{-{1\over 2h_{\G}^{\vee}}}$ with $X$ and $M$ kept fixed,
the Hamiltonian and spectral curve for the twisted
elliptic Calogero-Moser system with Lie algebra $\G$
reduce to the Hamiltonian and spectral curve for the
Toda system for the affine Lie algebra $(\G^{(1)})^{\vee}$.
This is the correct answer. Indeed,
in this limit, the gauge theory with adjoint hypermultiplet
reduces to the pure Yang-Mills theory,
and the Seiberg-Witten spectral curves for
pure Yang-Mills with gauge algebra $\G$ have been shown by
Martinec and Warner [35] to be
the spectral curves of the Toda system
for $(\G^{(1)})^{\vee}$.

\medskip
$\bullet$ The equations $R(k,z)=det(kI-L(z))$
for the spectral curves of $\G=D_n$
can be written explicitly in the trigonometric limit
$\tau\to i\infty$, in terms of a polynomial
$H(A)\equiv\prod_{j=1}^n(A^2-k_j^2)$ similar to (6.4)
$$
R(k,z)=
{m^2+mA-2k{m\over Z}\over m^2+2mA}H(A)
+
{mA+2k{m\over Z}\over m^2+2mA}H(A+m),\eqno(6.10)
$$
where we have introduced the more convenient
spectral parameter $Z$ by ${1\over Z}={1\over 2}{\rm coth}{z\over 2}$,
and the variable $A$ is defined by the quadratic relation
$$
A^2+mA+2k{m\over Z}-k^2=0.
$$
The effective prepotential can be evaluated explicitly
in the case of $\G=D_n$ for $n\leq 5$. Its logarithmic singularity
does reproduce the logarithmic singularities expected
from field theory considerations.

\medskip
$\bullet$ As in the known correspondences between Seiberg-Witten theory
and integrable models [5,30], we expect the following equation to hold
$$
{\partial \F\over\partial\tau}=H_{\G}^{twisted}(x,p),
\eqno(6.11)
$$
to hold. Note that the left hand side can be
interpreted in the gauge theory as a renormalization group equation.

\medskip
$\bullet$ For simply-laced $\G$, the curves $R(k,z)=0$ are modular
invariant. Physically, the gauge theories for these Lie algebras
are self-dual. For non simply-laced $\G$,
the modular group is broken to the congruence subgroup
$\Gamma_0(2)$ for $\G=B_n,C_n$, $F_4$, and to $\Gamma_0(3)$
for $G_2$. The Hamiltonians of the twisted Calogero-Moser systems
for non-simply laced $\G$ are also transformed under Landen
transformations into the Hamiltonians of the twisted Calogero-Moser system
for the dual algebra $\G^{\vee}$. It would be interesting to determine whether
such transformations exist for the
spectral curves or the corresponding gauge theories themselves.

\medskip

Spectral curves for certain gauge theories with classical gauge algebras and
matter in the adjoint representation have also been proposed in [15]
(see in particular
the papers by Witten,
Uranga, and Yokono),
based on branes in string theory and M-theory. Connections between branes
configurations associated with $\N=2$ gauge theories and integrable systems
have been put forward in [36].

\bigskip
\bigbreak

\centerline{{\bf ACKNOWLEDGMENTS}}

\bigskip

E. D. acknowledges the warm hospitality and generous support from
the Yukawa Institute for Theoretical Physics, from Ryu Sasaki and Takeo Inami 
at the ``Workshop on Gauge Theory and Integrable Models" at Kyoto, January 
1999, and from Tohru Eguchi and Norisuke Sakai at ``Supersymmetry and Unified 
Theory of Elementary Particles" at Kyoto, February 1999. 

\bigskip
\bigbreak

\centerline{\bf REFERENCES}
\bigskip

\item{[1]} N. Seiberg and E. Witten,
``Electro-magnetic duality, monopole condensation,
and confinement in $\N=2$ supersymmetric Yang-Mills theory",
Nucl. Phys. {\bf B 426} (1994) 19-53, hep-th/9407087;\hfil\break
N. Seiberg and E. witten,
``Monopoles, duality, and chiral symmetry breaking
in $\N=2$ supersymmetric QCD",
Nucl. Phys. {\bf B 431} (1994) 494, hep-th/9410167.

\item{[2]} A. Gorskii, I.M. Krichever, A. Marshakov, A. Mironov and A. Morozov,
``Integrability and Seiberg-Witten exact solution", Phys. Lett. {\bf B355}
(1995) 466, hep-th/9505035.

\item{[3]} R. Donagi and E. Witten,
``Supersymmetric Yang-Mills and integrable systems",
Nucl. Phys. {\bf B 460} (1996) 288-334, hep-th/9510101.

\item{[4]} A. Gorsky and N. Nekrasov, ``Elliptic Calogero-Moser System
from two-dimensional current algebra", hep-th/9401021;\hfil\break
N. Nekrasov,
``Holomorphic bundles and many-body systems", Comm. Math. Phys. {\bf 180}
(1996) 587;\hfil\break
E. Martinec, ``Integrable structures in supersymmetric gauge
and string theory", hep-th/9510204.

\item{[5]} E. D'Hoker and D.H. Phong,
``Calogero-Moser systems in $SU(N)$ Seiberg-Witten theory",
Nucl. Phys. {\bf B 513} (1998) 405-444, hep-th/9709053.

\item{[6]} M.A. Olshanetsky and A.M. Perelomov,
``Completely integrable Hamiltonian systems
connected with semisimple Lie algebras",
Inventiones Math. {\bf 37} (1976) 93-108;\hfil\break
M.A. Olshanetsky and A.M. Perelomov,
``Classical integrable finite-dimensional
systems related to Lie algebras",
Phys. Rep. {\bf 71 C} (1981) 313-400.

\item{[7]} E. D'Hoker and D.H. Phong,
``Calogero-Moser Lax pairs with spectral parameter
for general Lie algebras", Nucl. Phys. {\bf B 530} (1998)
537-610, hep-th/9804124.

\item{[8]} E. D'Hoker and D.H. Phong,
``Calogero-Moser and Toda systems for twisted and
untwisted affine Lie algebras",
Nucl. Phys. {\bf B 530} (1998) 611-640, hep-th/9804125.

\item{[9]} E. D'Hoker and D.H. Phong,
``Spectral curves for super Yang-Mills with adjoint
hypermultiplet for general Lie algebras",
Nucl. Phys. {\bf B 534} (1998) 697-719, hep-th/9804126.

\item{[10]} E. D'Hoker and D.H. Phong,
``Lax Pairs and Spectral Curves for Calogero-Moser and Spin
Calogero-Moser Systems", hep-th/9903002, to appear in {\it Regular and Chaotic 
Dynamics};\hfil\break
E. D'Hoker and D.H. Phong, 
``Seiberg-Witten Theory and 
Integrable Systems", hep-th/9903068, to appear in
{\it Proceedings of the Edinburgh Conference
on Seiberg-Witten and Whitham theories}, ed. by
H. Braden and I.M. Krichever.

\item{[11]}   W. Lerche, 
``Introduction to Seiberg-Witten theory and its stringy origins",
Proceedings of the {\it Spring School and Workshop
on String Theory}, ICTP, Trieste (1996),
hep-th/9611190, Nucl. Phys. Proc. Suppl. {\bf B 55} (1997) 83;\hfil\break
Y. Ohta, ``Non-perturbative solutions
in N=2 supersymmetric Yang-Mills theories: progress
and perspective", hep-th/9903182.

\item{[12]} I.M. Krichever and D.H. Phong,
``Symplectic forms in the theory of solitons",
hep-th/9708170, to appear in Surveys in Differential Geometry, Vol. III
\hfil\break
I.M. Krichever and D.H. Phong,
``On the integrable geometry of $N=2$ supersymmetric gauge
theories and soliton equations",
J. Differential Geometry {\bf 45} (1997) 349-389.

\item{[13]} R. Donagi, ``Seiberg-Witten integrable models", alg-geom/9705010;
\hfil\break
R. Carroll, ``Prepotentials and Riemann surfaces", hep-th/9802130;\hfil\break
R. Carroll, ``Remarks on Whitham and RG", hep-th/9712110,\hfil\break
D. Freed, ``Special K\"ahler manifolds", hep-th/9712042

\item{[14]} A. Marshakov,
``On integrable systems and supersymmetric gauge
theories", Theor. Math. Phys. {\bf 112} (1997)
791-826, hep-th/9702083.
A. Marshakov, ``Seiberg-Witten curves and integrable systems",
hep-th/9903252;\hfil\break
A. Mironov, ``WDVV equations and Seiberg-Witten theories",
hep-th/9903088.

\item{[15]} E. Witten, ``Solutions
of four-dimensional field theories via M Theory",
Nucl. Phys. {\bf B 500} (1997) 3, hep-th/9703166;\hfil\break
A. Brandhuber, J. Sonnenschein, S. Theisen,
and S. Yankielowicz,
``M-Theory and Seiberg-Witten curves: orthogonal
and symplectic groups",
Nucl. Phys. {\bf B 504} (1997) 175, hep-th/9705232;
\hfil\break
A. Giveon and D. Kutasov,
``Brane dynamics and gauge theory",
hep-th/9802067
K. Landsteiner, E. Lopez, and D. Lowe,
``New curves from branes", Nucl. Phys. {\bf B 516}
(1998) 273, hep-th/9708118;\hfil\break
A.M. Uranga, 
``Towards mass deformed $\N=4$ $SO(N)$ and $Sp(K)$ gauge
theories from brane configurations",
Nucl. Phys. {\bf B 526} (1998) 241-277, hep-th/9803054;\hfil\break
T. Yokono, 
``Orientifold four plane in brane configurations and $\N=4$ $USp(2N)$
and $SO(2N)$ theory",
Nucl. Phys. {\bf B 532} (1998) 210-226, hep-th/9803123;\hfil\break
K. Landsteiner, E. Lopez, and D. Lowe,
``Supersymmetric gauge theories from branes and orientifold
six-planes", hep-th/9805158.

\item{[16]} S. Kachru and C. Vafa,
``Exact results for N=2 compactifications of heterotic strings",
Nucl. Phys. {\bf B 450} (1995) 69,
hep-th/9505105;
\hfil\break
M. Bershadsky, K. Intriligator,
S. Kachru, D. Morrison, V. Sadov,
and C. Vafa,
``Geometric singularities and enhanced
gauge symmetries",
Nucl. Phys. {\bf B 481} (1996)
215, hep-th/9605200;
\hfil\break
S. Katz, A. Klemm, and C. Vafa,
``Geometric engineering
of quantum field theories",
Nucl. Phys. {\bf B 497} (1997) 173,
hep-th/9609239;
\hfil\break
S. Katz, P. Mayr, and C. Vafa,
``Mirror symmetry and exact solutions of 4D
$\N=2$ gauge theories", Adv. Theor. Math. Phys.
{\bf 1} (1998) 53, hep-th/9706110

\item{[17]} I.P. Ennes, S.G. Naculich,
H. Rhedin, and H. Schnitzer,
``One-instanton predictions of a Seiberg-Witten curve
from M-Theory:
the symmetric representation",
Nucl. Phys. {\bf B 533} (1998) 275-302,
hep-th/9804105;\hfil\break
I.P. Ennes, S.G. Naculich,
H. Rhedin, and H. Schnitzer,
``One-instanton predictions of a Seiberg-Witten curve
from M-Theory:
the anti-symmetric representation",
Int. J. Mod. Phys. {\bf A 14} (1999) 301-321;
hep-th/9804151;\hfil\break
I.P. Ennes, S.G. Naculich,
H. Rhedin, and H. Schnitzer,
``One-instanton predictions for non-hyperelliptic curves
derived from M-Theory",
Nucl.Phys. {\bf B536} (1998) 245-257, 
hep-th/9806144;\hfil\break
I.P. Ennes, S.G. Naculich,
H. Rhedin, and H. Schnitzer,
``One-instanton predictions of Seiberg-Witten curves
for product groups", hep-th/9901124;\hfil\break
I.P. Ennes, S.G. Naculich,
H. Rhedin, and H. Schnitzer,
``Two antisymmetric hypermultiplets
in N=2 SU(N) gauge theory: Seiberg-Witten curve
and M Theory interpretation",
hep-th/9904078.

\item{[18]} I.M. Krichever,
``The $\tau$-function of the universal Whitham hierarchy,
matrix models, and topological field theories",
Comm. Pure Appl. Math. {\bf 47} (1994) 437-475;
\hfil\break
T. Eguchi and S.K. Yang,
``Prepotentials of N=2 supersymmetric gauge theories
and soliton equations",
hep-th/9510183;\hfil\break
T. Nakatsu and K. Takasaki,
``Whitham-Toda hierarchy and N=2 supersymmetric
Yang-Mills theory",
Mod. Phys. Lett. {\bf A 11} (1996) 157-168,
hep-th/9509162;\hfil\break
K. Takasaki,
``Whitham deformations and tau functions
in N=2 supersymmetric gauge theories",
hep-th/9905224;\hfil\break
J. Edelstein, M. Gomez-Reina,
and J. Mas,
``Instanton corrections in N=2 supersymmetric theories
with classical gauge groups and fundamental
matter hypermultiplets",
hep-th/9904087.

\item{[19]} G. Bonelli and M. Matone,
``Nonperturbative relations in N=2 SUSY Yang-Mills and
WDVV equations",
Phys. Rev. Lett. {\bf 77} (1996) 4712,
hep-th/9606090;
\hfil\break
A. Marshakov, A. Mironov,
and A. Morozov,
``WDVV-like equations in N=2 SUST Yang-Mills theory",
Phys. Lett. {\bf B 389} (1996) 43, hep-th/9607109;
\hfil\break
J.M. Isidro,
``On the WDVV equation and M-Theory",
Nucl. Phys. {\bf B 539} (1999) 379-402.

\item{[20]} F. Calogero,
``Exactly solvable one-dimensional many-body problems",
Lett. Nuovo Cim. {\bf 13} (1975) 411-416.

\item{[21]} J. Moser, 
``Three integrable Hamiltonian systems connected with
isospectral deformations",
Advances Math. {\bf 16} (1975) 197.

\item{[22]} H. Braden, ``A conjectured R-matrix",
J. Phys. A {\bf 31} (1998) 1733-1741.

\item{[23]} A.P. Polychronakos,
``Generalized Calogero-Sutherland systems
from many matrix mo\-dels",
Nucl. Phys. {\bf B 546} (1999) 495-502;
\hfil\break
A.P. Polychronakos, ``Generalized Statistics in one dimension",
hep-th/9902157.

\item{[24]} I.M. Krichever, 
``Elliptic solutions of the Kadomtsev-Petviashvili equation
and integrable systems of particles",
Funct. Anal. Appl. {\bf 14} (1980) 282-290.

\item{[25]} H.W. Braden and V.M. Buchstaber,
``The general analytic solution
of a functional equation of addition type",
Siam J. Math. anal. {\bf 28} (1997) 903-923.

\item{[26]} A. Bordner, E. Corrigan, and R. Sasaki,
``Calogero-Moser systems: a new formulation", hep-th/9805106;\hfil\break
A. Bordner, R. Sasaki, and K. Takasaki,
``Calogero-Moser systems II: symmetries and foldings", hep-th/9809068;
\hfil\break
A. Bordner and R. Sasaki, ``Calogero-Moser
systems III: Elliptic potentials and twisting", hep-th/9812232;
\hfil\break
A. Bordner, E. Corrigan, and R. Sasaki,
``Generalized Calogero-Moser models and
universal Lax pair operators",
hep-th/9905011.

\item{[27]} I. Inozemtsev, 
``Lax representation with spectral parameter on a torus for integrable
particle systems", Lett. Math. Phys. {\bf 17} (1989) 11-17;
\hfil\break
I. Inozemtsev,
``The finite Toda lattices", Comm. Math. Phys. {\bf 121} (1989) 628-638.

\item{[28]} P. Goddard and D. Olive,
``Kac-Moody and Virasoro algebras in relation to
quantum physics", International J. Mod. Phys. A, Vol. I (1986)
303-414.

\item{[29]} E. D'Hoker and D.H. Phong,
``Order parameters, free fermions, and
conservation laws for Calogero-Moser systems",
hep-th/9808156, to appear in Asian J. Math.

\item{[30]} E. D'Hoker, I.M. Krichever, and D.H. Phong,
``The renormalization group equation for $\N=2$
supersymmetric gauge theories",
Nucl. Phys. {\bf B 494} (1997) 89-104, hep-th/9610156.

\item{[31]} E. D'Hoker, I.M. Krichever, and D.H. Phong, 
``The effective prepotential for $\N=2$ supersymmetric $SU(N_c)$ gauge
theories", Nucl. Phys. {\bf B489} (1997) 179, hep-th/9609041;
\hfil\break
E. D'Hoker, I.M. Krichever, and D.H. Phong,
``The effective prepotential for $\N=2$ supersymmetric
$SO(N_c)$ and $Sp(N_c)$ gauge theories",
Nucl. Phys. {\bf B 489} (1997) 211-222,
hep-th/9609145\hfil\break
E. D'Hoker and D.H. Phong, ``Strong Coupling Expansions of SU(N) Seiberg-Witten
Theory", Phys. Lett. {\bf B397} (1997) 94; hep-th/9701055.

\item{[32]} J. Minahan, D. Nemeschansky, and N. Warner,
``Instanton expansions for mass deformed $\N=4$ super Yang-Mills
theory", hep-th/9710146.

\item{[33]} M. Marino and G. Moore, ``The Donaldson-Witten Function for 
Gauge Groups of rank larger than one", hep-th/9802185;\hfil\break
M. Marino, ``The Uses of Whitham Hierarchies", hep-th/9905053.

\item{[34]} K. Vaninsky,
``On explicit parametrization of spectral curves
for Moser-Calogero particles and its applications",
December 1998 preprint.

\item{[35]} E. Martinec and N. Warner,
``Integrable systems and supersymmetric gauge theories",
Nucl. Phys. {\bf B 459} (1996) 97-112, hep-th/9509161.

\item{[36]} A. Gorsky, ``Branes and Integrability in the $\N=2$ SUSY
YM theory", Int. J. Mod. Phys. {\bf A12} (1997) 1243, hep-th/9612238;
\hfil\break
A. Gorsky, S. Gukov, A. Mironov, ``SUSY field theories, integrable 
systems and their stringy brane origin", hep-th/9710239;
\hfil\break
A. Cherkis and A. Kapustin, ``Singular monopoles and supersymmetric
gauge theories in three dimensions", hep-th/9711145.

\end